\documentclass[aps,prl,twocolumn,reprint,preprintnumbers,floatfix,nofootinbib]{revtex4-1}

\usepackage{amsmath}
\usepackage{amssymb}

\usepackage{graphicx}
\usepackage{subcaption}
\usepackage{dcolumn}
\usepackage{bm}
\usepackage{marvosym}
\usepackage{epstopdf}

\usepackage{color}
\definecolor{light-gray}{gray}{0.5}
\definecolor{blue}{rgb}{0.0,0.0,1.0}
\definecolor{green}{rgb}{0.0,0.5,0.0}
\definecolor{red}{rgb}{1.0,0.0,0.0}
\definecolor{cyan}{rgb}{0.0,0.75,0.75}
\definecolor{magenta}{rgb}{0.75,0.0,0.75}
\definecolor{yellow}{rgb}{0.75,0.75,0.0}

\newcommand{\avg}[1]{\langle{#1}\rangle}
\newcommand{\sdot}{\cdot}

\newcommand{\grad}{\bm \nabla}
\newcommand{\pd}{\partial}

\begin{document}
\title{Statistical equilibria of large scales in dissipative hydrodynamic turbulence}
\author{V. Dallas}
\author{S. Fauve}
\author{A. Alexakis}
\affiliation{Laboratoire de Physique Statistique, \'Ecole Normale Sup\'erieure, CNRS, Universit\'e Pierre et Mari\'e Curie, Universit\'e Paris Diderot, 24 rue Lhomond, 75005 Paris, France}

\begin{abstract}
We present a numerical study of the statistical properties of three-dimensional dissipative turbulent flows at scales larger than the forcing scale. Our results indicate that the large scale flow can be described to a large degree by the truncated Euler equations with the predictions of the zero flux solutions given by absolute equilibrium theory, both for helical and non-helical flows. Thus, the functional shape of the large scale spectra can be predicted provided that scales sufficiently larger than the forcing length scale but also sufficiently smaller than the box size are examined. Deviations from the predictions of absolute equilibrium are discussed.
\end{abstract}

\maketitle

%
Experimental and numerical studies of three-dimensional homogeneous hydrodynamic turbulent flows have been so far mostly focused on the finite energy flux solutions of the Navier-Stokes that manifest themselves on scales smaller that the forcing scale for which the Kolmogorov cascade and intermittency take place \cite{frisch95}. 
This is because the flows of many experiments designed to study statistically stationary turbulent regimes are forced at scales not much smaller than the size of the container. This is also the case of most direct numerical simulations (DNS) for which the flow is often forced in the largest possible modes 
aiming for the largest scale separation between the forcing scale and the small scales in the dissipative range.
A notable exception is of course the limit of two-dimensional flows for which the inverse cascade of energy \cite{kraichnan73} leads to a negative flux of energy that excites scales larger than the forcing scale. 

Many flows of geophysical or astrophysical interest far from the two-dimensional limit involve spatial structures at scales larger than the forcing scale. At these scales, no energy flux is expected and the usual Kolmogorov cascade picture does not hold. This is also true for some flows involved in industrial processes, such as large scale turbulent mixing. Dynamical and statistical properties of the zero flux solutions in scales larger than the forcing scale could thus be of interest for many applications in three-dimensional hydrodynamic turbulence.

Despite the lack of quantitative studies of the large scales in three-dimensional statistically stationary turbulence, it has been believed since a long time that the scales larger than the forcing scale
are in statistical equilibrium (see page 209 of reference \cite{frisch95}). The argument is that the energy driving the flow is transferred from the forcing scale $\ell_f$ to the dissipation scale $\ell_{\eta}$ by the Kolmogorov cascade and that no mean energy flux exists toward scales larger than $\ell_f$. The scales between $\ell_f$ and the container size $L$, thus do not involve any mean energy flux and could be in statistical equilibrium.

With this assumption a $k^2$ energy spectrum similar to the Rayleigh-Jeans spectrum for blackbody radiation would result with all modes in the range $2\pi /L < k < 2\pi /\ell_f$ being in equipartition. Such a 
spectrum has been obtained long ago using the Hopf equation for flows without forcing and viscosity \cite{hopf1952}. It is also the spectrum obtained in the absence of mean helicity in the truncated Euler equations (i.e. the Euler equations where only Fourier modes with wavenumbers $|{\bm k}| \le k_{cut}$ have been kept, $k_{cut}$ being the truncation wavenumber) \cite{orszag1977}. It should be noted that the steady state problem considered here differs from the one of the large scale structure in decaying turbulence, although a similar spectrum has been predicted \cite{saffman1967}.

When the initial conditions involve mean helicity $H$ in addition to kinetic energy $E$, both quadratic invariants need to be taken into account in deriving the energy and helicity distribution among scales for the truncated Euler system.
Following the statistical mechanics approach that is usually used to predict absolute equilibria of ideal homogeneous turbulence \cite{rosesulem78,shebalinNASA02} the Boltzmann-Gibbs equilibrium distribution is defined as $\mathcal P = Z^{-1} \exp(-\alpha E -\beta H)$, where $Z = \int_\Gamma \exp(-\alpha E -\beta H) d\Gamma$ is the partition function integrated over the phase space $\Gamma$ and $\alpha$, $\beta$ can be seen as the inverse temperatures in the classical thermodynamic equilibrium sense, which are determined by the total energy and the helicity of the system. From there Kraichnan \cite{kraichnan73} derived the absolute equilibria 
of the energy spectrum $E(k)$ and the helicity spectrum $H(k)$ which are
\begin{equation}
 E(k) = \frac{4\pi\alpha k^2}{\alpha^2 - \beta^2 k^2} 
\quad \text{and} \quad 
 H(k) = \frac{8\pi\beta k^4}{\alpha^2 - \beta^2 k^2}
 \label{eq:statmech}
\end{equation}
with $\alpha > 0$ and $\alpha > |\beta| k_{cut}$. These spectra have a singularity at $k = k_s \equiv \alpha/\beta > k_{cut}$ outside the range of validity of Eqs. \eqref{eq:statmech}. The ratio $|\beta| k_{cut}/\alpha$ gives a measure of the relative helicity $H(k)/(kE(k))$ of the flow with 0 corresponding to a non-helical flow, and 1 to the fully-helical singular case where all energy and helicity is concentrated in the largest wavenumbers $|{\bm k}| = k_{cut}$. 
The truncated Euler equations have been widely studied by Brachet and coworkers \cite{cichowlas2005} and the validity of the predicted spectra in Eqs. \eqref{eq:statmech} has been verified. 
A recent work has also shown that the kinematic dynamo properties of an ABC flow forced at small scales compared to the domain size can be well described by modelling the large scales of the flow using the truncated
Euler equation \cite{prasath2014}.

In this letter we show that despite the fact that in three-dimensional hydrodynamic turbulence the scales between the forcing scale and the container size are not isolated from the turbulent scales, their
statistics may still be reasonably approximated as if they were in statistical equilibrium.
We consider flows with high enough scale separation by applying helical and non-helical forcings at intermediate scales using numerical simulations of the forced hyperviscous Navier-Stokes equations and we focus on the dynamical and statistical properties of the large scales.

In laboratory experiments as well as in planets and stars, physical boundaries confine fluids and determine the largest possible length scales. In our DNS, the 
computational domain is the surrogate for this spatial confinement. For our study it is important to obtain high enough scale separation between the size of 
our periodic box $2\pi$ and the forcing scale while at the same time small scale turbulence is resolved.
Forcing at intermediate scales and aiming for a turbulent flow with high enough scale separation is almost prohibitive even with today's supercomputing power. We partly circumvent this difficulty by considering the hyperviscous Navier-Stokes equations under the assumption that the viscous scale should not significantly affect the statistical properties of the large scales. The hyperviscous Navier-Stokes equations then read as
\begin{equation}
 \pd_t \bm u + (\bm u \sdot \grad) \bm u = - \grad P + (-1)^{h+1} \nu_h \grad^{2n} \bm u + \bm f
 \label{eq:hyperNS}
\end{equation}
where $\bm u(\bm x,t)$ denotes the solenoidal velocity field, $\nu_h$ is the specified constant hyperviscocity, $\bm f$ is the forcing function, which is described below and $P$ is the hydrodynamic pressure. Note that for our purposes the hyperviscous term was chosen to take the value of $n = 4$.
In the ideal 
case $\nu_h = 0$ and ${\bm f=0}$ Eq. \eqref{eq:hyperNS} conserves the kinetic energy $E = \frac{1}{2}\avg{|\bm u|^2}$ and the helicity $H = \avg{\bm u \cdot \bm \omega}$ with $\bm \omega = \grad \times \bm u$ being the vorticity and angular brackets denoting a spatial average unless indicated otherwise. The level of helicity in the flow corresponds to the degree of the alignment between the velocity and the vorticity and this is given by the normalized helicity $-1 \leq \rho_H \equiv H/(\avg{|\bm u|^2}\avg{|\bm \omega|^2})^{1/2} \leq 1$.

Using a standard pseudo-spectral code we numerically solve Eq. \eqref{eq:hyperNS} 
satisfying $\grad \cdot \bm u = 0$. Aliasing errors are removed using the $2/3$ 
rule, i.e. wavenumbers $k_{min} = 1$ and 
$k_{max} = N/3$, where $N$ is the number of grid points 
on each side of the computational box. The temporal integration was performed using a third-order Runge-Kutta scheme. Further details on the code can be found in \cite{mpicode05}.

In this study, the velocity field is forced at intermediate wavenumbers $k_f$. The forcing that we consider are a helical random forcing
 \begin{align}
 \bm f_H = f_0
  \{&[\cos(k_f y + \phi_y) + \sin(k_f z + \phi_z)]\hat x, \nonumber \\
    &[\cos(k_f z + \phi_z) + \sin(k_f x + \phi_x)]\hat y, \nonumber \\
    &[\cos(k_f x + \phi_x) + \sin(k_f y + \phi_y)]\hat z\} 
\label{eq:f_H}
 \end{align}
where $\bm f_H \cdot \grad \times \bm f_H = k_f\bm f_H^2 > 0$ at each point in space and a non-helical random forcing
 \begin{align}
 \bm f_{NH} = f_0
  \{&[\sin(k_f y + \phi_y) + \sin(k_f z + \phi_z)]\hat x, \nonumber \\
    &[\sin(k_f z + \phi_z) + \sin(k_f x + \phi_x)]\hat y, \nonumber \\
    &[\sin(k_f x + \phi_x) + \sin(k_f y + \phi_y)]\hat z\}
 \end{align}
where $\avg{\bm f_{NH} \cdot \grad \times \bm f_{NH}} = 0$. The phases $\phi_x$, $\phi_y$, $\phi_z$ were changed randomly at given correlation time scales $\tau_c$. All the necessary parameters of our problem are tabulated below (see Table \ref{tbl:dnsparam}). Here we define the Reynolds number based on our control parameters as $Re \equiv u_f k_f^{1-2n}/\nu_h$, where $u_f\propto (f_0/k_f)^{1/2}$.

\begin{table}[!ht]
  \caption{Numerical parameters of the DNS. Note that 
	   $\tau_f \equiv (k_{min}f_0)^{-1/2}$.}
  \label{tbl:dnsparam}
    \begin{tabular}{c|*{6}{c}}
     \hline
     \hline
     $\,\,\, k_f \,\,\,$ & $\,\,\, \rho_H \,\,\,$ & $\,\,\, f_0 \,\,\,$ & $\,\,\, \tau_c/\tau_f \,\,\, $ & $\,\,\, \nu_h\,\,\,$ & $Re$ & $N$ \\
    \hline
      10 & 0.6 & 1.0 & 0.3   & $5 \times 10^{-12}$ & $6.3 \times 10^{3}$  &128 \\
      20 & 0.6 & 2.0 & 0.15  & $5 \times 10^{-15}$ & $3.5 \times 10^{4}$  &256 \\
      40 & 0.6 & 4.0 & 0.075 & $1 \times 10^{-17}$ & $9.7 \times 10^{4}$  &512 \\
      10 & 0.0 & 1.0 & 0.3   & $5 \times 10^{-12}$ & $6.3 \times 10^{3}$  &128 \\
      20 & 0.0 & 2.0 & 0.15  & $5 \times 10^{-15}$ & $3.5 \times 10^{4}$  &256 \\
      40 & 0.0 & 4.0 & 0.075 & $1 \times 10^{-17}$ & $9.7 \times 10^{4}$  &512 \\
     \hline
     \hline
   \end{tabular}
\end{table}

\begin{figure}[!ht]
 \begin{subfigure}{0.23\textwidth}
   \includegraphics[width=\textwidth]{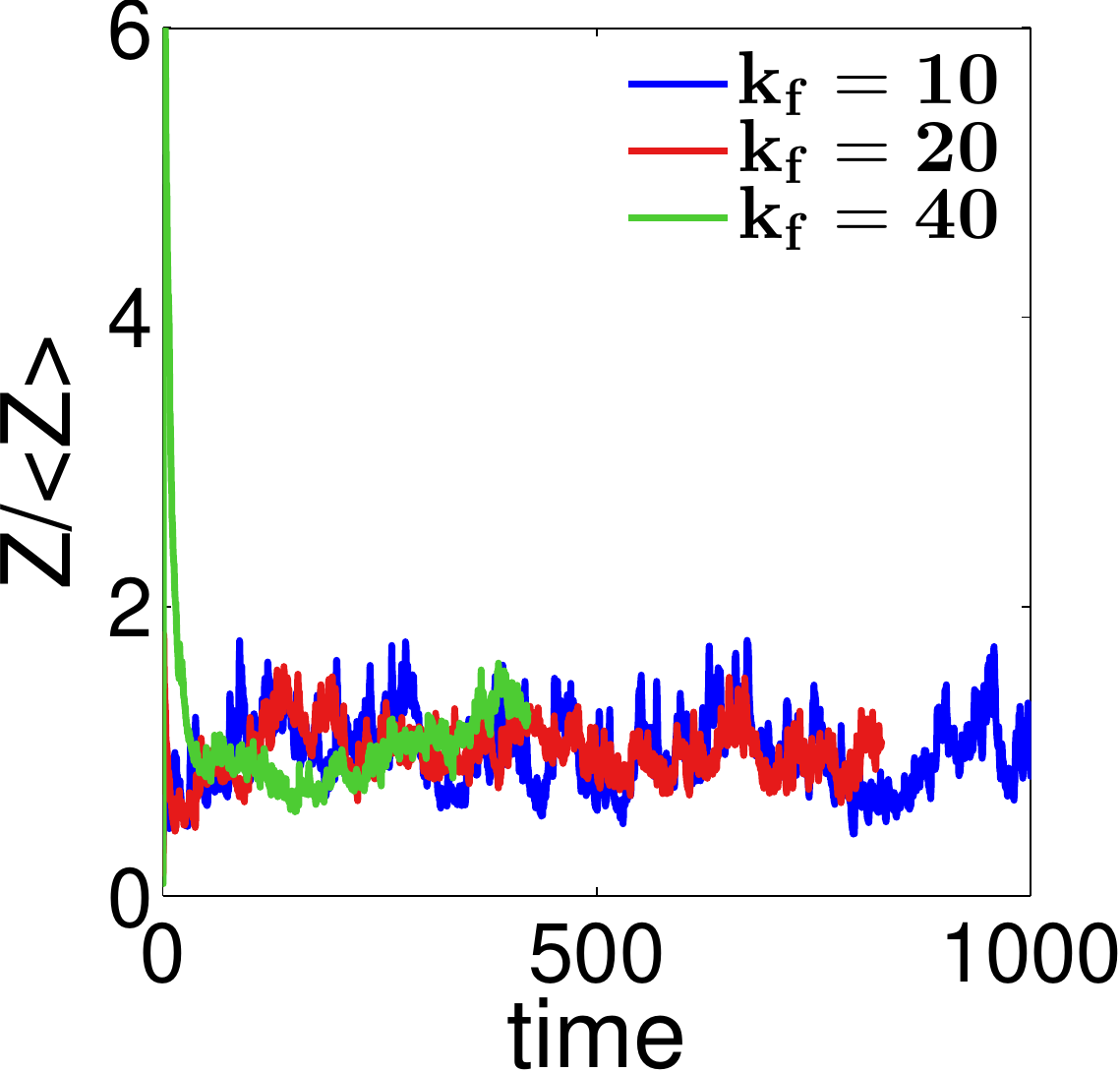}
   \caption{}
   \label{fig:sata}
 \end{subfigure}
 \begin{subfigure}{0.23\textwidth}
   \includegraphics[width=\textwidth]{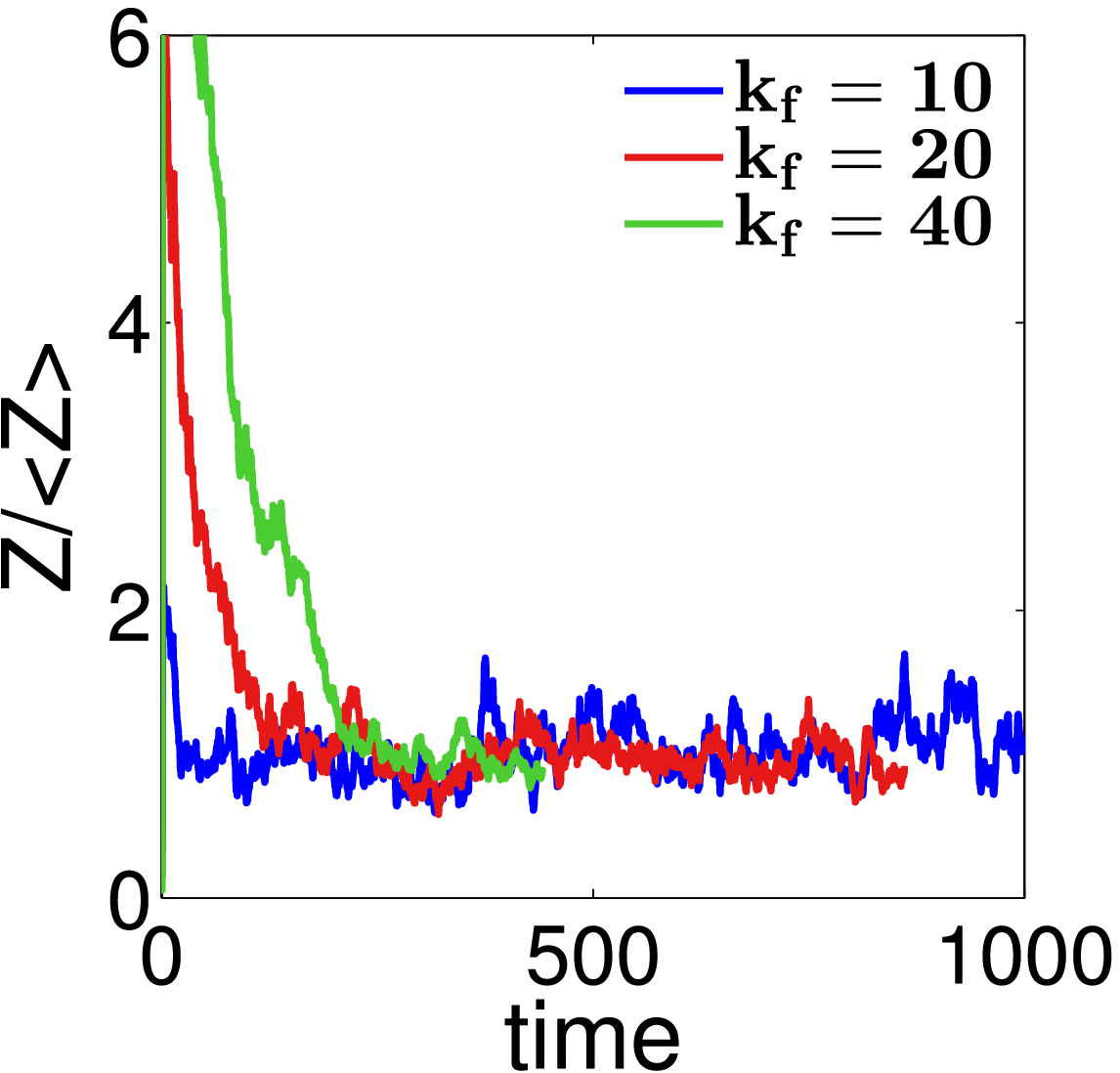}
   \caption{}
   \label{fig:satb}
 \end{subfigure}
  \caption{(Color online) Large scale quantity $Z$ normalised by its time-average $\avg{Z}$ as a function of time for (a) helical and (b) non-helical flows.}
  \label{fig:saturation}
 \end{figure}

Since we are interested on the large scale behavior we need to make sure that our DNS have been integrated long enough so that the largest scales are in a statistically stationary state. In order to illustrate that such states have been reached, we define the energy weighted in the large scales as $Z(t) = \sum_{k,k\ne0} \ k^{-4}E(k,t)$. $Z$ is a large scale quantity and we monitor it as a function of time (see Fig. \ref{fig:saturation}).
After long enough time integration the large scales reach a stationary state for both helical (Fig. \ref{fig:sata}) and non-helical flows (Fig. \ref{fig:satb}). 
In what follows we analyze the data from this saturated states.

\begin{figure}[!ht]
 \begin{subfigure}{0.49\textwidth}
  \includegraphics[width=\textwidth]{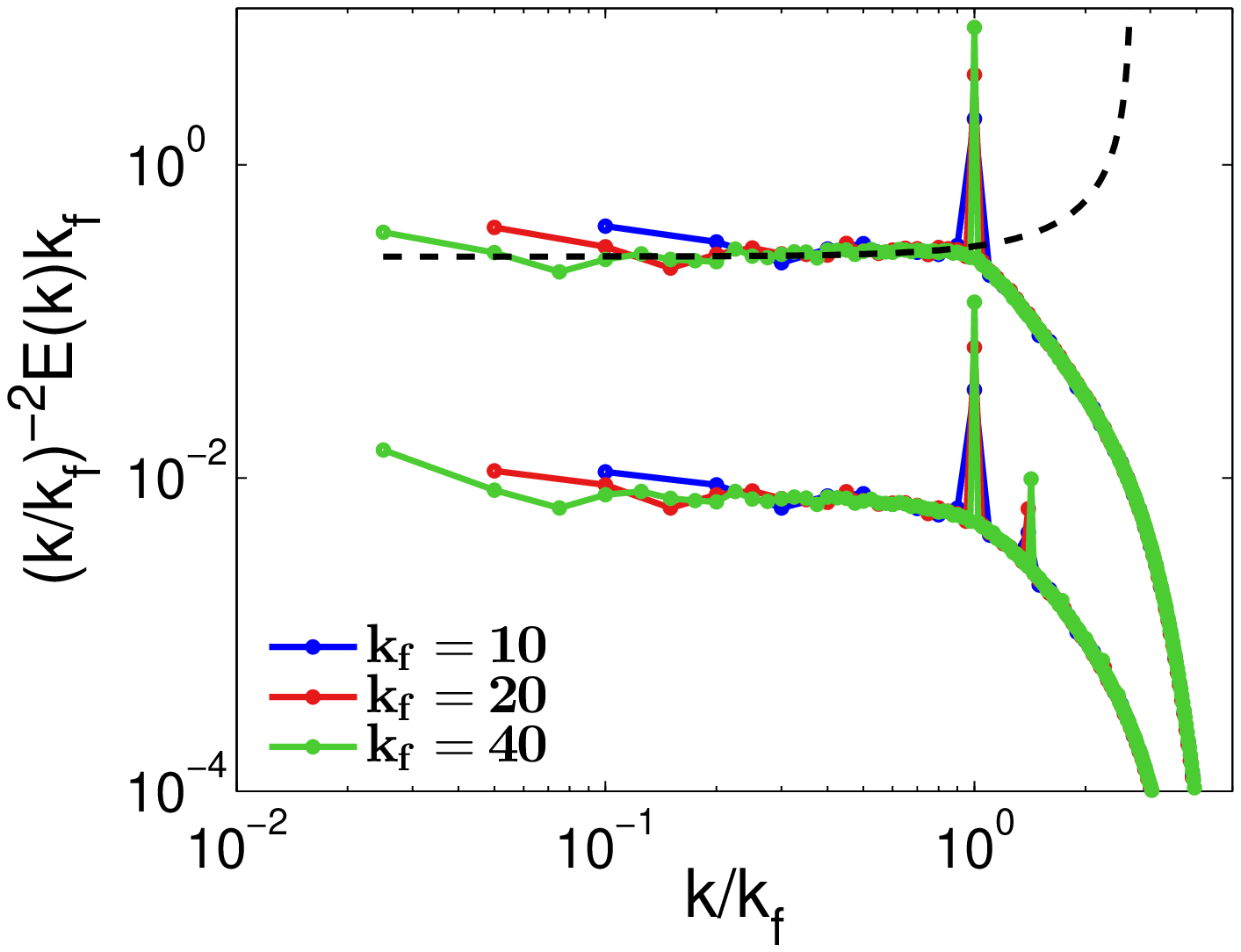}
   \caption{}
   \label{fig:Espec}
 \end{subfigure}
 \begin{subfigure}{0.49\textwidth}
   \includegraphics[width=\textwidth]{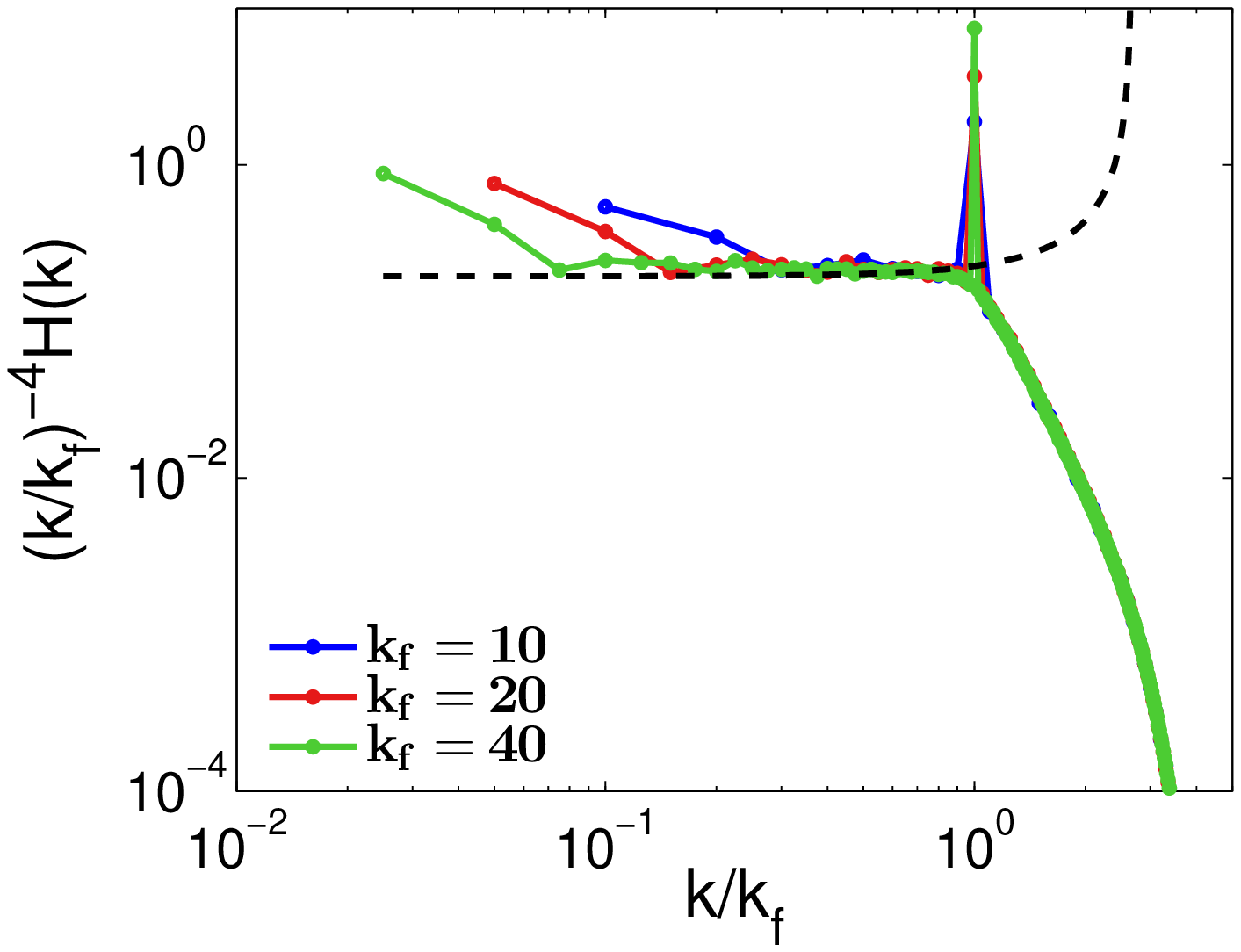}
   \caption{}
   \label{fig:Hspec}
 \end{subfigure}
  \caption{(Color online) (a) Compensated $k^{-2}E(k)$ energy spectra for helical (top) and non-helical (bottom) flows. (b) Compensated $k^{-4}H(k)$ helicity spectra. The dotted lines represent Kraihnan's absolute equilibria (Eqs. \eqref{eq:statmech}).}
  \label{fig:LSspec}
 \end{figure}

Figure \ref{fig:LSspec} presents the energy spectra compensated with $k^{-2}$ (Fig. \ref{fig:Espec}) and the helicity spectra compensated with $k^{-4}$ (Fig. \ref{fig:Hspec}). Note that the energy and helicity spectra collapse since they are rescaled with $k/k_f$. In Fig. \ref{fig:Espec} the energy spectra for the helical and non-helical flows are shown with the non-helical spectra being shifted down for clarity. 
Our data displays a $E(k) \propto k^2$ scaling at low wavenumbers $k < k_f$ both for the helical and the non-helical flows. Similarly the collapsed helicity spectra in Fig. \ref{fig:Hspec} display the scaling $H(k) \propto k^4$. These scalings are in agreement with the absolute equilibria of the truncated Euler equations for helical and non-helical flows. 
For comparison to the Kraichnan's theory, we have plotted Eqs. \eqref{eq:statmech} as dotted lines (see Fig. \ref{fig:LSspec}) 
using values of $\alpha$ and $\beta$ obtained from a linear fit. 
These curves indicate that the divergence of the spectra predicted by Kraichnan's Eqs. \eqref{eq:statmech} at $k_s = \alpha/\beta$ is expected at $k_s \simeq 2.5 k_f$ which is well beyond the expected validity of the absolute equilibrium regime. For this reason no singular behavior is observed deviating from the $H(k) \propto k^4$ scaling and the $E(k) \propto k^2$ power law due to the presence of helicity.

\begin{figure}[!ht]
  \includegraphics[width=0.49\textwidth]{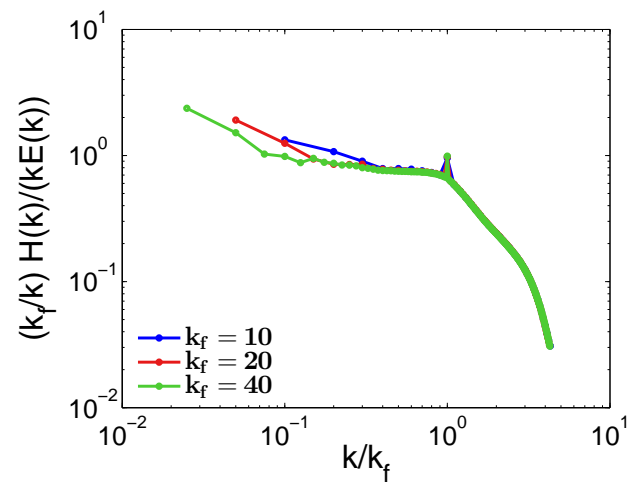}
   \caption{Relative helicity spectra $H(k)/(kE(k))$ rescaled with $k/k_f$.}
   \label{fig:H/E}
 \end{figure}

To investigate the effect of helicity in the large scales for the helical runs we plot the relative helicity spectra rescaled with $k/k_f$ in Fig. \ref{fig:H/E}. Kraichnan's absolute statistical equilibria 
(Eqs. \eqref{eq:statmech}) imply that $(k_f/k)H(k)/(kE(k))$ is equal to the non-dimensional number $2\beta k_f/\alpha = 2k_f/k_s$.
This ratio appears to be approximately constant
for the highest $k_f$ runs and only for the range of wavenumbers $3k_{min} \leq k < k_f$. The measured value of this ratio in this range gives $2\beta k_f/\alpha \simeq 0.8$ indicating the amount of the relative helicity in the large scales. Despite the fully helical forcing (Eq. \ref{eq:f_H}) used, not enough helicity has been transfered in the large scales to make the flow fully helical (i.e. $\beta k_f/\alpha = 1$). 

Deviations from Eqs. \eqref{eq:statmech} do exist at the largest scales of the system $k \leq 2k_{min}$. These scales appear to be more energetic and more helical than absolute equilibrium predicts. There are many possible reasons for this behavior.
First for modes with wavelengths close to the box size the assumptions of isotropy used in the derivation of Eqs. \eqref{eq:statmech} are not valid and deviations from the isotropic result are expected. 
Another possibility is that a large scale instability could be present \cite{AKA}. Such an instability can transfer energy directly from the forced and turbulent scales to the largest scale of the flow
and alter the distribution of energy among modes in the steady state. 

\begin{figure}[!ht]
  \begin{subfigure}{0.23\textwidth}
  \includegraphics[width=\textwidth]{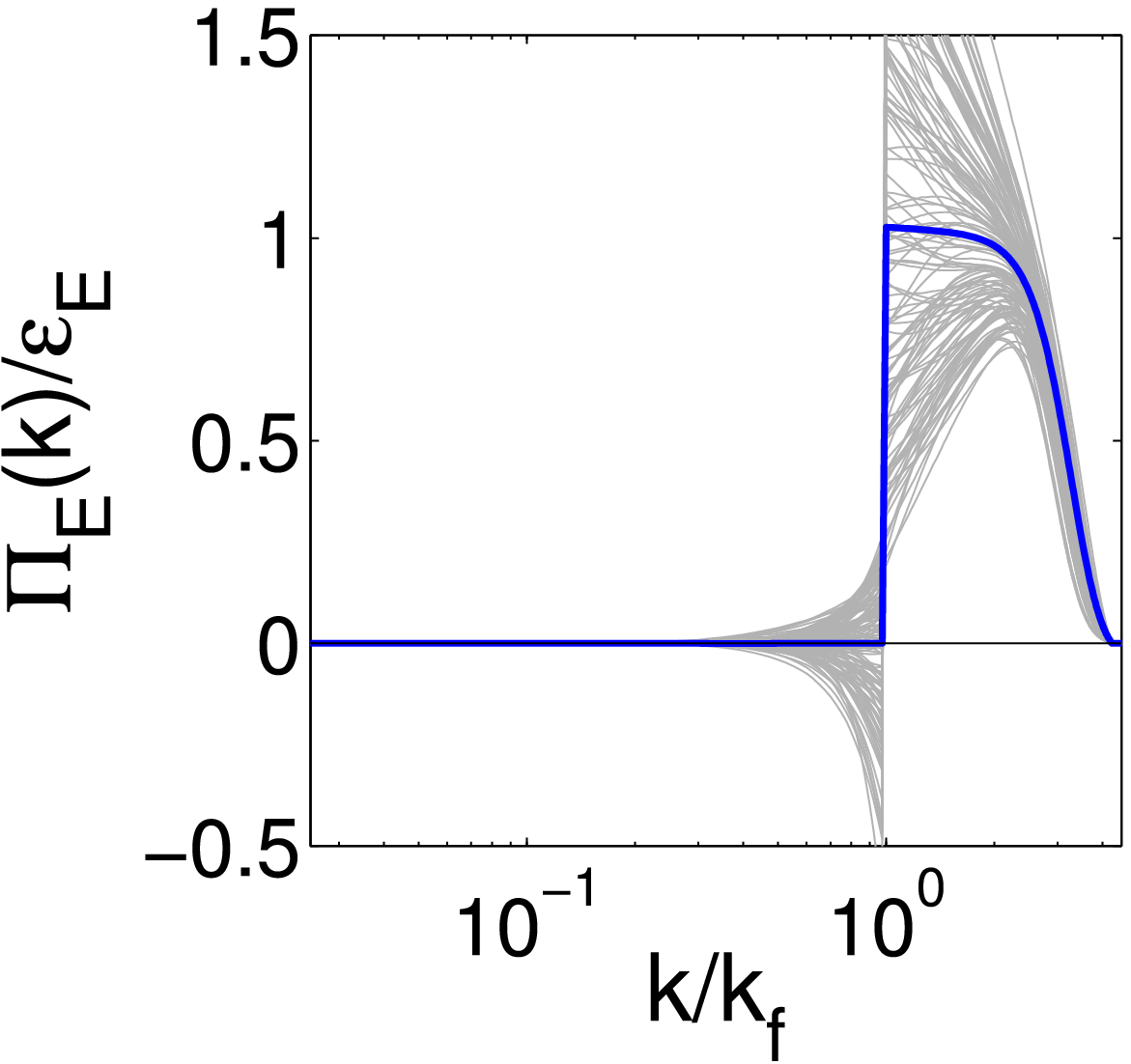}
   \caption{}
   \label{fig:fluxE}
  \end{subfigure}
  \begin{subfigure}{0.23\textwidth}
  \includegraphics[width=\textwidth]{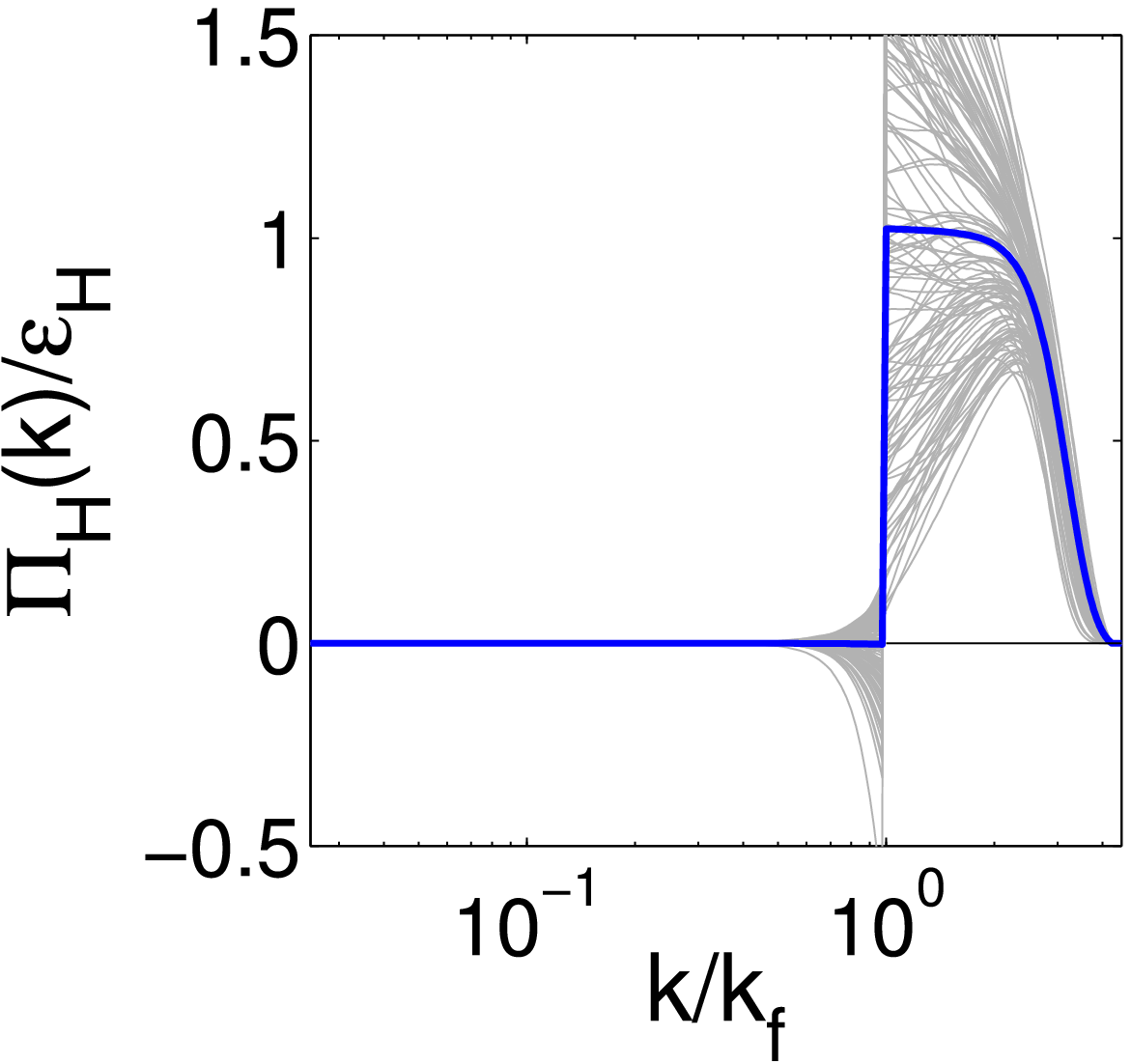}
   \caption{}
   \label{fig:fluxH}
  \end{subfigure}
  \caption{(Color online) (a) $\Pi_E(k)/\epsilon_E$ spectra and (b) $\Pi_H(k)/\epsilon_H$ spectra for the helical flow with $k_f=40$. Thick blue lines represent the time-averaged values while thin gray lines the instantaneous values for various instants in time.}
  \label{fig:flux}
 \end{figure}

%
At steady state no inverse cascade (negative flux) is expected in three-dimensional hydrodynamic turbulence for either energy or helicity. Figure \ref{fig:flux} shows the energy flux $\Pi_E(k)$ (Fig. \ref{fig:fluxE}) and the helicity flux $\Pi_H(k)$ (Fig. \ref{fig:fluxH}) normalised by the energy dissipation rate $\epsilon_E = 2\nu \int_0^{\infty} k^2 E(k,t) dk$ and the helicity dissipation rate $\epsilon_H = 2\nu \int_0^{\infty} k^2 H(k,t) dk$ respectively for the helical flow with $k_f=40$.
For the wavenumbers $k > k_f$ both fluxes are positive and constant over the range of $k_f < k < 2.5k_f$ signifying a forward energy and helicity cascade. In the $k < k_f$ range both time averaged fluxes are zero as expected for absolute equilibria. However, even though the time-averaged $\Pi_E(k)$ and $\Pi_H(k)$ are zero this is not true for the instantaneous fluxes that have large fluctuations of both signs.
These fluctuations imply a transfer of energy and helicity towards and from the turbulent scales in such a way that on average the total flux is zero. 
This exchange of energy with the turbulent small scales is in disagreement with the assumptions of absolute equilibrium that the modes following the spectra in Eqs. \eqref{eq:statmech} are isolated from external sources and sinks of energy.


In this letter we investigated to what extend the large scale flow in three-dimensional dissipative hydrodynamic turbulence can be described by the absolute statistical equilibria exhibited from the truncated Euler equations. Using numerical simulations we focus at the spectra of the energy and helicity at large scales. We considered both helical and non-helical flows which were forced at intermediate wavenumbers. For the non-helical flows we observed a $k^2$ energy spectrum at large scales, where the energy is equally distributed among the wavenumbers $k < k_f$. For the helical flows a $k^2$ energy spectrum persisted at large scales and the helicity spectrum displayed a $k^4$ power law at $k < k_f$ in agreement to Kraichnan's theory for ideal helical flows \cite{kraichnan73}. 

Despite the fully helical forcing used not enough helicity was transfered in the large scales to allow us to test the singularity of the spectra at $k_s = \alpha / \beta$ that would also distinguish the scaling of the energy spectra between the helical and the non-helical flows. In absolute equilibria of flows without forcing and dissipation of energy the values of the inverse temperatures are determined by the initial conditions. However, in this dissipative system is not clear how the system selects these values. 

A measurable deviation in the energy and helicity spectra was also observed at the largest scales of the system. Scales of size similar to the box were observed to be more helical and more energetic than the absolute equilibrium predictions. We speculate that these deviations are either due to the absence of isotropy in these scales or due to the presence of a large scale instability.

Energy and helicity fluxes were also investigated. The energy and helicity have a forward cascade for $k > k_f$ and no cascade (zero-flux) for $k < k_f$. Notably, even though the time-averaged energy and helicity flux is zero, the absolute equilibrium spectra at large scales are formed and sustained by flux fluctuations of the energy and the helicity. The presence of these fluctuations implies that there is energy exchange between the large scale flow and the turbulent small scale fluctuations. Whether these fluctuations play a sub-dominant role or whether they provide a different mechanism for the formation of the $k^2$ spectra is a question that requires further investigation.

To conlude, the present results provide support to the relevance of the absolute equilibrium spectra to the behavior of the large scales in forced dissipative turbulent flows despite the fact that scales between the forcing scale and the domain size ($k < k_f$) are not isolated from the turbulent small scales ($k > k_f$).

\begin{acknowledgements}
The authors aknowledge enlightening discussions with M. E. Brachet. VD acknowledges the financial support from the Association DEPHY (Project No. 814443E). The computations were performed using the HPC resources from GENCI-TGCC-CURIE (Project No. x2014056421).
\end{acknowledgements}
\bibliography{references}
\end{document}